\documentclass{Interspeech2024}
\usepackage{amsmath}
\usepackage{flushend}
\usepackage{balance}
\usepackage{subcaption}
\usepackage{graphicx}

\interspeechcameraready

\title{MakeSinger: A Semi-Supervised Training Method for Data-Efficient Singing Voice Synthesis via Classifier-free Diffusion Guidance}
\name[affiliation={1}]{Semin}{Kim}
\name[affiliation={1}]{Myeonghun}{Jeong}
\name[affiliation={1}]{Hyeonseung}{Lee}
\name[affiliation={1}]{Minchan}{Kim}
\name[affiliation={1}]{Byoung Jin}{Choi}
\name[affiliation={1}]{Nam Soo}{Kim}

\address{
  $^1$Department of Electrical and Computer Engineering and INMC, \\Seoul National University, Seoul, South Korea}
\email{\{smkim21, mhjeong, hslee, bjchoi, mckim\}@hi.snu.ac.kr, nkim@snu.ac.kr}


\begin{document}
\maketitle

\begin{abstract}
    
    In this paper, we propose MakeSinger, a semi-supervised training method for singing voice synthesis~(SVS) via classifier-free diffusion guidance. The challenge in SVS lies in the costly process of gathering aligned sets of text, pitch, and audio data. MakeSinger enables the training of the diffusion-based SVS model from any speech and singing voice data regardless of its labeling, thereby enhancing the quality of generated voices with large amount of unlabeled data. At inference, our novel dual guiding mechanism gives text and pitch guidance on the reverse diffusion step by estimating the score of masked input. Experimental results show that the model trained in a semi-supervised manner outperforms other baselines trained only on the labeled data in terms of pronunciation, pitch accuracy and overall quality. Furthermore, we demonstrate that by adding Text-to-Speech~(TTS) data in training, the model can synthesize the singing voices of TTS speakers even without their singing voices.
\end{abstract}

\section{Introduction}

Singing voice synthesis~(SVS) is a rapidly growing field of research adopting various deep learning techniques. Similar to Text-to-Speech~(TTS), a typical two-stage pipeline of neural SVS consists of an acoustic model and a vocoder. The acoustic model \cite{blaauw2020sequence, lee2021n, hono2021sinsy, tae2021mlp} generates acoustic features such as mel-spectrogram from the musical score, including text, tempo, and pitch. The vocoder \cite{45774, yamamoto2020parallel, kong2020hifi, huang2021multi} then converts these acoustic features to a waveform. Recent advances include end-to-end models that aim to improve the sound quality and naturalness such that the generated audio results in a more human-like voice \cite{Lee2019AdversariallyTE, chen2020hifisinger, zhang2022visinger, zhang2022visinger2}.

However, training such models demands a high-quality dataset with precisely aligned (\textit{text}, \textit{pitch}, \textit{audio}) triplets. This makes collecting data for SVS challenging, as it involves professional singers who can accurately sing to the pitch and tempo of the musical score. Semi-supervised approaches have emerged to address this issue, leveraging extracted pseudo labels from unlabeled data.
For instance, \cite{choi2022melody} employed Harvest \cite{morise2017harvest} to generate pitch labels, and phoneme classifier for text labels. DeepSinger \cite{ren2020deepsinger} gathered unlabeled data from publicly available music websites while extracting the pitch using Parselmouth\footnote{\url{https://github.com/YannickJadoul/Parselmouth}}.
Despite their effectiveness, these approaches come with a significant limitation: they are not entirely free from data labeling. They still require (\textit{text}, \textit{audio}) pair for the training data, or require an auxiliary classifier specifically tailored for singing voices, which eventually demands a large amount of labeled data to prevent potential error propagation from inaccurate labeling.

Meanwhile, diffusion generative models \cite{anderson1982reverse, ho2020denoising, song2020score} have shown significant breakthroughs in many areas of deep learning. Notably, studies such as \cite{popov2021grad, liu2022diffsinger, jeong2021diff} have demonstrated the remarkable effectiveness of diffusion models in TTS and SVS. Additionally, \cite{kim2021guided, kim2022guided} introduced a training method that capitalizes on the strengths of diffusion models, incorporating the training of a separate phoneme classifier to provide text guidance. To avoid building the classifier, \cite{ho2021classifier} introduced classifier-free guidance, enabling a model to learn from both conditional and unconditional data and generate class-conditional samples by a single joint model. These guiding approaches facilitate efficient learning from unlabeled datasets.

Inspired by these works, we propose MakeSinger, a semi-supervised training method for data-efficient singing voice synthesis that leverages any speech or singing voice, regardless of its labeling. MakeSinger employs classifier-free guidance, taking advantage of the joint model's capability to learn from both labeled and unlabeled data. We also present a novel dual guiding mechanism that adjusts the balance between text and pitch guidance during reverse diffusion steps. Our experimental results indicate that incorporating a substantial amount of unlabeled singing voice data into training enhances the pronunciation and overall quality of the synthesized voices, outperforming baselines supervised with more labeled songs. Furthermore, by leveraging the multi-speaker TTS dataset, we demonstrate MakeSinger's ability to generate the singing voices of diverse TTS speakers, even without specific singing voice data for those speakers. We present our samples generated by MakeSinger on our demo page \footnote{\url{https://makesinger.github.io/MakeSinger-demo}}.

\section{Proposed Method}\label{section:Method}

 \begin{figure*}[ht]
 \centering
 \begin{subfigure}{0.47\linewidth}
 
 \raggedright
 \includegraphics[width=.9\linewidth]{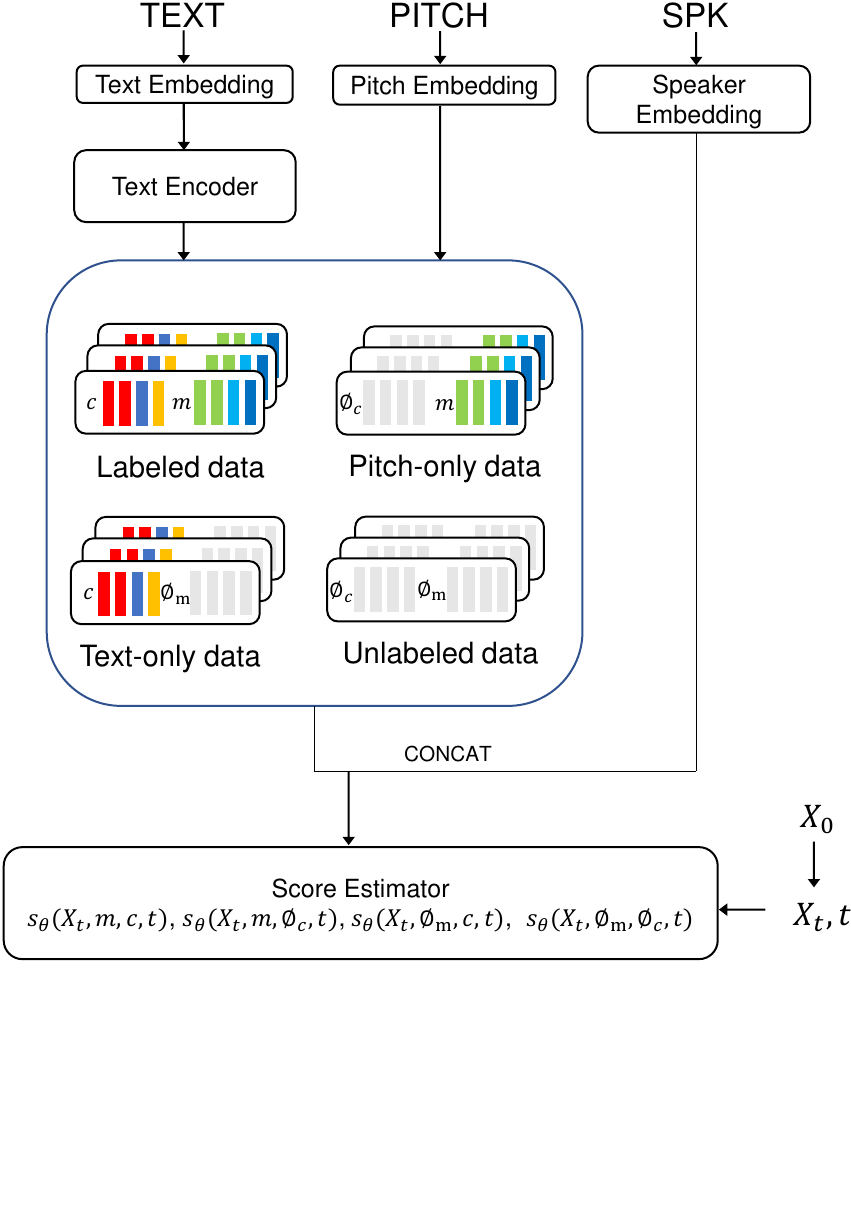}
 \\
 \caption{Training Procedure}
 \label{fig:A}
 \end{subfigure}
 \begin{subfigure}{0.47\linewidth}
 \raggedleft
 \includegraphics[width=.9\linewidth]{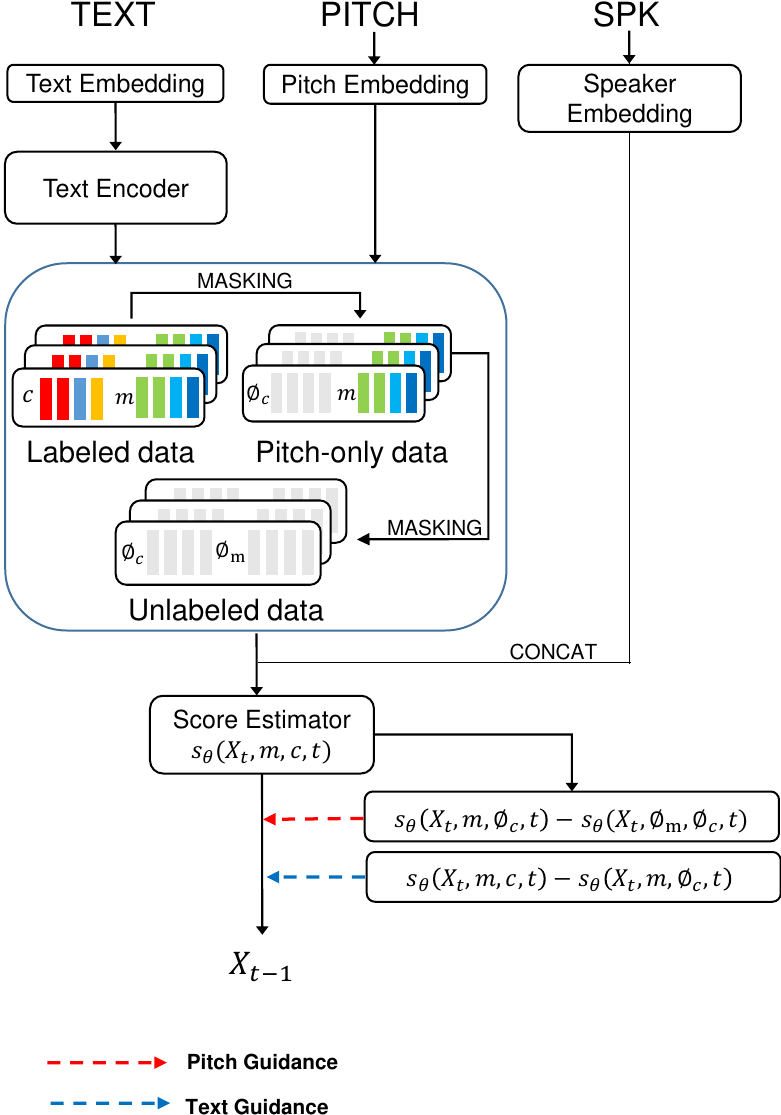}
 \\
 \caption{Inference Procedure}
 \label{fig:B}
 \end{subfigure}
 \caption{ Overall procedure of MakeSinger.}
 \label{fig:fig1}
 \end{figure*}

In this section, we describe MakeSinger, our proposed semi-supervised training method for SVS. As a base architecture, we use Grad-TTS \cite{popov2021grad}, a score-based TTS model, with several modifications. In training procedure, our diffusion model learns from any audio without relying on its labeling, contributing the
robustness of the system in the low-resource scenario with a limited availability of paired data. During inference, we apply a dual guiding mechanism to provide a balanced and appropriate level of pitch and text guidance. We explain detailed architecture of MakeSinger in Section 2.1, and our dual guiding mechanism in Section 2.2. The comprehensive overview of the MakeSinger is illustrated in Figure \ref{fig:fig1}.




\subsection{Architecture}
MakeSinger is designed to generate a mel-spectrogram corresponding to the given text, pitch, and speaker. When the
model learns from labeled data, temporally aligned text and pitch sequence along with its mel-spectrogram becomes an input. In case where neither text nor pitch label exists, we use $\langle unknown \rangle$  token sequence as substituting label. MakeSinger consists of two main parts: encoder and score estimator.

\noindent \textbf{Encoder} :
The encoder part converts text, pitch, and speaker information into the feature sequence of mel-spectrogram frame
length. Text encoder is consists of a pre-net, 6 Transformers \cite{vaswani2017attention}, and a linear projection layer. All the feature sequences are concatenated together to give conditional information to the score estimator. As in Figure \ref{fig:fig1}, we denote the feature sequences extracted from the text and pitch as $c$ and $m$. We additionally denote the feature sequence extracted from  $\langle unknown \rangle$ text token as $\emptyset_c$ and $\langle unknown \rangle$ pitch token as $\emptyset_m$. Unlike in Grad-TTS, our encoder solely focuses on encoding the conditional information without estimating the mean and variance of the prior noise distribution, as the strong prior information can diminish the effect of the guiding process in Section \ref{dual}.

\noindent \textbf{Score estimator}:
The score estimator $s_\theta$, which has U-Net \cite{ronneberger2015u} architecture, predicts the score at the given diffused mel-spectrogram $X_t$ and the concatenated feature sequence, where $t$ stands for the diffusion timestep. When training the model, we first pick uniformly random $t$ between 0 and 1. Then, we transform the ground-truth mel-spectrogram $X_0$ to the diffused mel-spectrogram $X_t$ through a forward diffusion process. We leverage score-matching loss, an L2 loss between the predicted and computed scores of $X_t$ to train the score estimator.

\subsection{Dual guiding in diffusion generative model}
\label{dual}

Original diffusion guiding mechanism \cite{song2020score} employs an additional classifier that demands enough labeled data to train the classifier as in Guided-TTS \cite{kim2021guided}. To address the low resource problem of the singing voice, we adopt a classifier-free guiding from \cite{ho2021classifier} in a semi-supervised manner. This method involves providing guidance at each reverse diffusion step, facilitating class-conditional generation. \\ 
In our guiding scheme for generating singing voices, we recognize the necessity to manage underlying textual and pitch information. A simple approach is modeling the guidance for text and pitch with a single term. However, this approach hinders separate control over text and pitch, limiting the model's flexibility. To ensure finer control, we opt for dual guidance, applying distinct guidance terms for text and pitch. We formulate the dual guidance for diffused mel-spectrogram $X_t$, diffusion timestep $t$, and conditional information $y$. We start from a general diffusion guiding formulation from \cite{song2020score}:
%
\begin{align}
\nabla_{X_t}\log p_t(X_t|y) = \nabla_{X_t}\log p_t(X_t) + \nabla_{X_t}\log p_t(y|X_t).
\end{align}
We split $y$ to pitch condition $m$ and text condition $c$. Then, applying Bayes' rule, we can get classifier-free guidance for both labels:
\begin{flalign}
       &\nabla_{X_t}\log p_t(m, c|X_t) \nonumber\\
       &= \nabla_{X_t}\log p_t(X_t|m, c) - \nabla_{X_t} \log p_t(X_t)\nonumber \\
       &=(\nabla_{X_t}\log p_t(X_t|m, c) - \nabla_{X_t} \log p_t(X_t|m))\nonumber \\
       &+ (\nabla_{X_t} \log p_t(X_t|m) - \nabla_{X_t} \log p_t(X_t)).
\end{flalign}
Finally, we can derive the score using dual guiding with scaling factors:
\begin{flalign}\label{eq.3}
      s_t &= \nabla_{X_t}\log p_t(X_t|m,c) \nonumber \\
      &+ w_1(\nabla_{X_t}\log p_t(X_t|m, c) - \nabla_{X_t} \log p_t(X_t|m)) \nonumber  \\
      &+ w_2(\nabla_{X_t} \log p_t(X_t|m) - \nabla_{X_t} \log p_t(X_t)),
\end{flalign}
where $s_t$ is a time-dependent total score for the reverse time process, $w_1$ and $w_2$ are scaling factors for text and pitch guidance. 
By a similar approach, we can also derive the following:
\begin{flalign}\label{eq.4}
      s_t &= \nabla_{X_t}\log p_t(X_t|m,c) \nonumber \\
      &+ w_1(\nabla_{X_t}\log p_t(X_t|m, c) - \nabla_{X_t} \log p_t(X_t|c)) \nonumber  \\
      &+ w_2(\nabla_{X_t} \log p_t(X_t|c) - \nabla_{X_t} \log p_t(X_t)).
\end{flalign}
%
Note that we provide guidance to the score estimated from conditional input since it generates better samples compared to giving guidance to the unconditional score. 
Actual implementation for Eq.~\ref{eq.3} is expressed as follows:
\begin{flalign}\label{eq.5}
      s_t &= s_\theta(X_t, m, c, t) + \alpha^{1}_tw_1(s_\theta(X_t, m, c, t) \nonumber \\
      &- s_\theta(X_t, m, \emptyset_c, t)) + \alpha^{2}_tw_2(s_\theta(X_t, m, \emptyset_c, t) \nonumber\\
      &- s_\theta(X_t, \emptyset_m, \emptyset_c, t)).
\end{flalign}
%
We emphasize that a single joint model trained with both conditional and unconditional data estimates all the score terms. The model masks the original feature sequence to pitch-only and unlabeled data sequences to estimate each guidance, as depicted in Figure \ref{fig:B}. By employing norm-based guidance from \cite{kim2021guided}, we optimize the control by scaling each norm of the guiding gradient in proportion to the norm of the score of the joint model in every diffusion timestep $t$. $\alpha^{1}_t$ and $\alpha^{2}_t$ in Eq. \ref{eq.5} are normalizing factors of the norm-based guidance.


\section{Experimental Settings}
\subsection{Datasets}
We conducted multi-speaker SVS experiments on the the open-sourced Korean multi-speaker singing voice dataset \footnote{\url{https://bit.ly/3uZc7WM}}. We employed 1,210 ballad songs sung by 12 singers. We spitted it into 973 songs for training and 277 songs for the test. We also conducted SVS experiments utilizing TTS data from open-sourced large-scale Korean multi-speaker dataset \footnote{\url{https://bit.ly/4c46sz6}}, aiming to synthesis singing voices of TTS speakers. All audio files were resampled to 22,050 Hz and we computed the mel-spectrograms using a 1,024-point short-time Fourier transform with 256-point hop size and mel-scale filterbanks featuring 80 bins.

\subsection{Input Processing}
Our dataset is paired with musical instrument digital interface~(MIDI) files that serve as labels. As MIDI file contains aligned pitch and text expressed in syllables according to the tempo, we extracted $m$, $c$, $f^{s}$, and $f^{e}$ of length $I$ from each song. $m$, $c$ denotes sequence of pitches and Korean syllables respectively, and $f^{s}_{i}$, $f^{e}_{i}$ correspond to start and end frame index of $i$ th pitch and syllable. $f^{s}_i$, $f^{e}_i$ is computed by considering the frequency of waves and mel-spectrogram frame length. Since a Korean syllable can be divided into three phonemes: an onset, a nucleus, and an optional coda, we allocated frames for each phoneme. In Korean, most vocalization time spends on pronouncing vowels, which is always the nucleus. Following the approach of \cite{tae2021mlp, Lee2019AdversariallyTE, choi2020korean}, we allocated three frames for the onset and coda, the rest for the nucleus while dividing the frames of a syllable. \\
For TTS data, We used Montreal Forced Aligner~(MFA) \cite{mcauliffe2017montreal}, which automatically extracts speech-phoneme alignments, to extend syllables to mel-spectrogram length. In order to standardize the phoneme set for both the SVS and TTS, we applied the same allocation of onset, nucleus, and coda as we did for the singing voice data.
\subsection{Model Configuration}
We trained the following models for comparison in the multi-speaker SVS experiments. 
\begin{itemize}
    \item \textbf{MLP-Singer}: MLP-Mixer \cite{tolstikhin2021mlp} based Korean SVS model.
    \item \textbf{VISinger}: End-to-end SVS model based on VITS \cite{kim2021conditional}. Given that our experimental data were pre-aligned, we removed the duration predictor and length regulator from the original model. Additionally, we substituted its encoder to ours for fair comparison. 
    \item \textbf{MakeSinger without semi-supervision}: Our proposed model only supervised with labeled data.
    \item \textbf{MakeSinger with semi-supervision}: Our proposed model with semi-supervised training. Dual guiding was applied in the inference procedure.
\end{itemize}
To demonstrate the effectiveness of our proposed method and its correlation with the amount of labeled data, we varied the number of labeled songs across three different settings, maintaining constant 12 singers: 12, 36, and 108 songs. MLP-Singer, VISinger, and MakeSinger without semi-supervision exclusively utilized labeled songs for training. \\
For the MakeSinger with semi-supervision, we extended the training set by including the remaining unlabeled songs. Specifically, we used [12, 961], [36, 937], [108, 865] labeled and unlabeled songs. During inference, we employed Eq. 3 from Section 2 for reverse diffusion steps. To improve the gradient estimation of $\nabla_x \log p_t(X_t|m)$, we introduced pitch-only data in the training data created by masking text labels of the labeled data. We empirically configured the scaling factor for text guidance $w_1 = 0.2$ and pitch guidance $w_2 = 0.02$ in the dual guiding process.\\
Both MakeSinger models used 200 reverse diffusion steps and shared hyperparameter settings with Grad-TTS. Unlike \cite{song2020score, popov2021grad} that used an ordinary differential equation~(ODE) without the stochastic term in the inference process, we opted for a stochastic differential equation~(SDE) to better capture the vibrato commonly found at the end of sentences in ballad songs. \\
As MLP-Singer and MakeSinger generate mel-spectrogram, we used Hifi-GAN \cite{kong2020hifi} to generate the waveform. We fine-tuned its universal pre-trained version\footnote{\url{https://github.com/jik876/hifi-gan}} to our singing voice dataset since Hifi-GAN is originally trained to synthesis speech.
MLP-Singer is trained on a single 2080Ti GPU. All the other models were trained up to 60 epochs on 4 2080Ti GPUs.

\section{Results}
\subsection{Main Results}
To verify above four systems in three different settings, we performed a subjective evaluation using the mean opinion score~(MOS) test, and for objective evaluation, we utilized F0 root mean square error~(F0-RMSE) and semitone accuracy~(S-ACC). Results are presented in Table. \ref{tab:my-table}\\
\textbf{Subjective Evaluation Metrics}:
The MOS test involved 14 participants and was conducted with 95 \% confidence intervals. We evaluated 24 randomly sampled song segments each 4 to 6 seconds long, focusing on pronunciation accuracy (MOS-P) and overall quality (MOS-Q). For pronunciation accuracy, we instructed testers to evaluate how naturally the singer pronounced the text in terms of ballad singing. For overall quality, we instructed testers to evaluate the comprehensive quality of singing, including sound quality, pronunciation, pitch accuracy, and naturalness.\\
\textbf{Objective Evaluation Metrics}:
We computed F0-RMSE and S-ACC against reference audios reconstructed by a vocoder. S-ACC measures how accurately a system can determine the pitch of a musical note in semitones. Randomly selected 350 segments of songs between 4 to 6 seconds were used to compute average score. \\
\textbf{Results}:
As shown in Table.~\ref{tab:my-table}, even MakeSinger without semi-supervision outperformed other baseline systems for every metric when trained with 36 and 108 songs. Furthermore, MakeSinger with semi-supervision achieved best scores across all evaluations, demonstrating the effectiveness of dual guiding mechanism, especially in pitch accuracy. Remarkably, with just 12 or 36 labeled songs, the semi-supervised MakeSinger showed comparable or superior performance to other baseline models trained on 108 labeled songs in every metric, highlighting our model's data efficiency. This efficiency was most pronounced with the smallest labeled dataset, where the semi-supervised model showed significant improvement over its non-semi-supervised counterpart. The performance gap between the two proposed models becomes smaller as the quantity of labeled songs increased, which aligns with the expectation that the importance of guiding decreases as more labeled data becomes available for the model to learn from.
\begin{table}[th]
\setlength{\tabcolsep}{2pt}
\setlength{\arrayrulewidth}{0.2mm}
\caption{Experimental results in terms of MOS-P, MOS-Q, F0-RMSE, and S-ACC.}
\label{tab:multi}
\centering
\begin{tabular}{l c c c c}
\toprule
\textbf{Method} & \multicolumn{1}{c}{\textbf{MOS-P}}& \multicolumn{1}{c}{\textbf{MOS-Q}}& \multicolumn{1}{c}{\textbf{\begin{tabular}[c]{@{}c@{}}F0-RMSE\\ ($\downarrow$)\end{tabular}}}& \multicolumn{1}{c}{\textbf{\begin{tabular}[c]{@{}c@{}}S-ACC\\ ($\uparrow$)\end{tabular}}} \\ 
\midrule
Reconstruction        & 4.64$\pm$0.07 & 4.47$\pm$0.08 & - & -  \\
\midrule
\textbf{12 labeled songs}\\
MLP-singer      & 2.79$\pm$0.11 & 2.52$\pm$0.10 &0.2468&41.76                                  \\
VISinger      & 2.20$\pm$0.11 & 2.19$\pm$0.09 &0.3005&37.14                                  \\
MakeSinger \\
\textit{- w/o semi-sup.}        & 2.37$\pm$0.11 & 2.54$\pm$0.11 &0.2543&48.98                               \\
\textit{- w/ semi-sup.}      & \textbf{3.30$\pm$0.12} & \textbf{3.48$\pm$0.12}
&\textbf{0.1736}&\textbf{54.29}                                  \\
\midrule
\textbf{36 labeled songs}\\
MLP-singer      & 3.14$\pm$0.11 & 2.71$\pm$0.09 &0.2419&42.95                                  \\
VISinger      & 2.74$\pm$0.11 & 2.59$\pm$0.09 &0.2121&45.91                                  \\
MakeSinger \\
\textit{- w/o semi-sup.}      & 3.41$\pm$0.11 & 3.36$\pm$0.11 &0.1942&52.51                                   \\
\textit{- w/ semi-sup.}      & \textbf{3.61$\pm$0.12} & \textbf{3.66$\pm$0.12}
&\textbf{0.1720}&\textbf{54.25         }                        \\
\midrule
\textbf{108 labeled songs}\\
MLP-singer      & 3.45$\pm$0.11 & 3.01$\pm$0.09 &0.2233&46.06                                  \\
VISinger      & 3.51$\pm$0.11 & 3.33$\pm$0.10 &0.2059&50.98                                   \\
MakeSinger\\
\textit{- w/o semi-sup.}      & 3.95$\pm$0.10 & 3.76$\pm$0.09 &0.1971&52.57                                  \\
\textit{- w/ semi-sup.}      & \textbf{4.03$\pm$0.09} & \textbf{4.08$\pm$0.09}
&\textbf{0.1702}&\textbf{56.05}                                   \\
\bottomrule
\label{tab:my-table}
\end{tabular}
\end{table}


\subsection{Ablation study}
To validate our dual guiding mechanism, we carried out an ablation study on the same semi-supervised MakeSinger model, comparing three different approaches during inference: dual guiding, single guiding, and no guiding. Single guiding approach models text and pitch guidance in a single term, which is expressed as follows:
\begin{flalign}
      s_t &= s_\theta(X_t, m, c, t) + \alpha_t w(s_\theta(X_t, m, c, t) \nonumber \\
      &- s_\theta(X_t, \emptyset_m, \emptyset_c, t)).
\end{flalign}
We set $w = 0.2$ in the single guiding to match the scale of guidance. 
The result in Table 2 shows that the single guiding mechanism degraded MakeSinger's pronunciation and overall quality, whereas completely omitting the guiding mechanism led to further degradation.

\begin{table}[th]
\setlength{\tabcolsep}{2pt}
\setlength{\arrayrulewidth}{0.2mm}
\caption{Ablation study with MOS-P and MOS-Q.}
\label{tab:multi}
\centering
\begin{tabular}{l c c}
\toprule
\textbf{Method} & \multicolumn{1}{c}{\textbf{MOS-P}}& \multicolumn{1}{c}{\textbf{MOS-Q}}\\
\midrule
Reconstruction        & 4.54$\pm$0.10 & 4.49$\pm$0.10 \\
\midrule

\textbf{MakeSinger}\\
 \textit{- no guiding}      & 3.80$\pm$0.10 & 3.77$\pm$0.07                               \\
 \textit{- single guiding}     & 3.87$\pm$0.10 & 3.84$\pm$0.07                              \\
 \textit{- dual guiding}      & \textbf{3.95$\pm$0.10} & \textbf{3.94$\pm$0.06}                                \\
\bottomrule
\label{tab:my-table2}
\end{tabular}
\end{table}

\subsection{Singing Voice Synthesis based on TTS data}
Utilizing TTS data helps MakeSinger to generate singing voices in TTS speakers' voices while maintaining the characteristics and timbre of the TTS speaker. Exploiting the high quality and high coverage of multi-speaker TTS data can also be a breakthrough to the insufficient number of singers and songs in the SVS dataset. Eq. \ref{eq.4} is more suitable for the task, as the model learns to estimate $\nabla_x \log p_t(X_t|c)$ from the TTS data, which only has text labels. \\
To show the feasibility of generating singing voices of TTS speakers, we incorporated multi-speaker TTS data into the training process. This involved using 487 labeled songs along with the remaining unlabeled songs and 3 hours of TTS data uttered by 80 speakers. Results show that MakeSinger is capable of generating singing voices from TTS speakers with dependable quality, as presented in our demo page.

\section{Conclusion}
We presented MakeSinger, a diffusion-based semi-supervised training method for singing voice synthesis. Exploiting the advantage of the classifier-free guidance, we could utilize any human speech and singing voice data regardless of its labeling. Furthermore, during the reverse diffusion steps, our novel dual guiding mechanism provides better controllability to the model regarding text and pitch. To the best of our knowledge, MakeSinger is the first SVS model to leverage classifier-free diffusion guiding in a semi-supervised manner. By applying our method, we showed that the model can generate singing voices with better pronunciation and sound quality. Also, we demonstrated that MakeSinger generates songs in the voice of TTS speakers by utilizing multi-speaker TTS data in training.

\section{Acknowledgements}
This work was supported by Institute of Information $\&$ communications Technology Planning $\&$ Evaluation (IITP) grant funded by the Korea government(MSIT) (No.2021-0-00456, Development of Ultra-high Speech Quality  Technology for Remote Multi-speaker Conference System)

\bibliographystyle{IEEEtran}
\bibliography{mybib}

\end{document}